# Layer-by-layer growth and growth-mode transition of SrRuO$_3$ thin films on atomically flat single-terminated SrTiO$_3$ (111) surfaces

Jaewan Chang, Yoon-Seok Park, Jong-Woo Lee, and Sang-Koog Kim [a]

Research Center for Spin Dynamics & Spin-Wave Devices (ReC-SDSW) and

Nanospinics Laboratory, Department of Materials Science and Engineering, College of Engineering,

Seoul National University, Seoul 151-744, Republic of Korea

**Abstract**

We report on growth-mode transitions in the growth of SrRuO$_3$ thin films on atomically flat Ti$^{4+}$ single-terminated SrTiO$_3$ (111) substrates, investigated by reflection high-energy electron diffraction and atomic force microscopy. Over the first ~9 unit cells, the dominant growth mode changes from island to layer-by-layer for the growth rate of 0.074 unit cells/sec and the growth temperature of 700 ºC. Moreover, in the course of growing SrRuO$_3$ films, the governing growth mode of interest can be manipulated by changing the growth temperature and the growth rate, which change allows for the selection of the desired layer-by-layer mode. The present study thus paves the way for integrations of SrRuO$_3$ thin layers into (111)-orientated oxide heterostructures, and hence multi-functional devices, requiring control of the sharp atomic-level interfaces and the layer-by-layer growth mode.

[a] Author to whom all correspondence should be addressed. E-mail: sangkoog@snu.ac.kr

Controlling the crystallographic orientation of thin films by altering that of a substrate on which the films are grown has been one of the pursuits for fabrications of oxide heterostructures of new and better physical properties. For instance, a manipulation of a nearest neighbor atomic ordering through an alternate stacking of one unit-cell layers in two perovskite oxides on (111)-oriented substrates has been theoretically predicted to lead to a different magnetic ordered state, which would not be observed from bulk materials.[1] In addition, such ferroelectric oxide films as $SrBi_2Nb_2O_9$ and $BiFeO_3$, which would not be polar-axis oriented on (001)-oriented substrates, have exhibited enhanced spontaneous electric polarizations when they are grown on (111)-oriented substrates.[2,3]

Not only as a ferromagnetic functional component in (111)-oriented functional heteroepitaxial devices, but an electrode in (111)-oriented ferroelectric capacitors, free from fatigue and imprinting behaviors, $SrRuO_3$ (SRO) thin films would be one of promising candidates to be integrated into such (111)-oriented structures, as have been favorably integrated into (001)-oriented heterostructures.[4,5] The key to realizing (111)-oriented oxide heterosturctures is the control of sharp heterointerfaces on the atomic level, and thus the control of a favorable growth mode, the so-called layer-by-layer growth mode is prerequisite. Despite such a strong need for (111)-oriented oxide heterostructures, realizing the layer-by-layer growth of SRO films on $SrTiO_3$ (STO) (111) has yet to be reported because an atomically well-defined, single-terminated STO (111) surface has proved difficult to achieve. A recent accomplishment of successful fabrications of such a STO (111) surface by Chang *et al.*,[6] however, has opened the door to the layer-by-layer growth of SRO thin films, enabling SRO thin films to be integrated into (111)-oriented oxide heterostructures.

In this letter, we report on anomalous growth-mode transitions at the initial growth of SRO thin films on single-terminated STO (111) substrates. We found that the growth modes during the initial growth stages are controllable by adjusting the growth rate ($R_g$) and growth temperature ($T_g$). The results presented here reveal that SRO thin films can be grown in the layer-by-layer fashion over a wide thickness range, making possible sharp atomic-level interfaces in (111)-orientated oxide heterostructures.

SRO thin films were grown on $Ti^{4+}$ single-terminated STO (111) substrates by pulsed layer deposition (PLD). The specifics of the preparation of single-terminated STO (111) substrates were described in our previous paper.[6] For stoichometric ablation, a KrF excimer laser (wavelength λ=248 nm) was focused on a sintered SRO target with a fluence of 2.5 J/cm$^2$. The oxygen pressure was maintained at 100 mTorr. During the SRO film deposition, the growth modes were monitored by *in situ* high-pressure reflection high-energy electron diffraction (RHEED) [Pascal]. A 20 keV electron beam with an incidence angle of around 1° was directed along [1$\bar{1}$0] azimuth. The RHEED diffraction patterns and intensity profiles in specular reflection geometry were recorded using a charge-coupled device camera with an acquisition software (K-Space Associates). The surface morphologies of the grown SRO films were probed by atomic force microscopy (AFM).

Figure 1(a) shows the RHEED intensity variation during the growth of an SRO film at $T_g$ =700 ºC and $R_g$=0.074 unit cells/s. The growth rates of the grown films in this study were determined by temporal period of the RHEED intensity oscillations. Two different growth-mode regimes are distinguishable (see the dashed vertical line), as revealed by the two distinct characteristic RHEED profiles. The earlier growth regime (hereafter denoted as regime I) prior to the first ~9 unit cells, is governed by the 3D island mode, excepting the layer-by-layer growth up to the first three oscillations. The absence of RHEED-intensity oscillation, corresponding to the first unit-cell layer, might be attributed to a terminating layer conversion from $Ru^{4+}$ to $SrO_3^{4-}$ [ref. 7]. The 3D island mode in regime I is changed to the layer-by-layer mode after the first ~9 unit cells, that is, to the later growth regime, denoted as regime II. In order to avoid the roughening behavior and extend the desired layer-by-layer mode into the earlier growth regime, we increased the growth temperature to $T_g$=800 ºC for the same growth rate ($R_g$=0.074 unit cells/s). The RHEED intensity profile shown in Fig 1(b) reveals that the higher $T_g$ allows for the prevalence of the layer-by-layer mode up to regime I, but after further growth, above the first ~9 unit cells, an unidentified growth mode dominates, which is characterized not only by the damped amplitude of the RHEED intensity oscillation but also by the lack of intensity recovery.

The observed roughening behavior between the third and the ninth unit-cell layers, shown in Fig. 1, may not be understood in terms of kinetics. The abrupt growth-mode transitions and the reentrance of the layer-by-layer mode above the ninth unit-cell layers could be explained by an alternative possible mechanism, i.e., an electrostatic polarity effect and its compensation. This effect is associated with the polar nature of the constituent planes of the films. As previously reported,[8,9] growing an ionic material along a polar plane leads to a surface roughening to suppress the increase of an electric potential with film thickness. A few monolayer thick SRO film can be an insulator due to its low dimensionality, and as the thickness of the SRO film is greater than ~1.98 nm [corresponding to ~9 unit-cell layers in SRO/STO (111)], the film shows a metallic behavior, as found in Ref. 10. Accordingly, in the thickness regime showing insulating properties, the SRO film would exhibit a rough surface above the third unit-cell layer, while the film would bear the increase in the potential up to the third unit-cell layer as have been predicted in MgO and ZnO ultrathin (111) films.[11] In the thickness regime showing metallic behaviors, however, free charge carriers, originating from the recovered metallicity at the thickness of ~9 unit-cell layers would compensate the electric field inherent to the SRO films, and the relevant electrostatic polarity effect does not affect the roughening of the SRO film morphology any more.

To verify those growth modes identified from the RHEED intensity profiles, the surface morphologies of the grown SRO films were measured by AFM after interrupting the deposition at the stages marked by the open and closed arrows within regimes I and II, respectively. Figure 2(a) shows the AFM images of the SRO thin films grown at $T_g$=700 ºC and $R_g$=0.074 unit cells/s, after the interruptions of the deposition within regimes I and II denoted in Fig. 1(a). The AFM image for regime I shows many islands of 1~2 nm height and ~25 nm lateral size, which islands are randomly distributed on the step-and-terrace of the STO (111) substrate. The diffused streaks and the additional streaks (enclosed by white circles) in the corresponding RHEED pattern (left top) evidence the presence of crystalline disorders and transmission through the 3D islands, respectively. The AFM image corresponding to regime II, by contrast, shows the surface morphology of the fully covered terraces and steps of $d_{111}$ (the interplanar spacing of (111) planes) height, as well as sharp spots of the RHEED pattern, indicating that the layer-by-layer growth mode prevails in regime II. The remarkable contrasts in the RHEED intensity profiles and the AFM surface morphologies between the 3D island and layer-by-layer growth modes [see Figs. 1(a) and 2(a)] evidence that the thickness-dependent growth mode transition takes place from the 3D island to layer-by-layer mode during the initial growth stages of SRO on STO (111).

Figure 2(b) shows the AFM surface morphologies of SRO thin films grown at the higher temperature, $T_g$=800 ºC, for the same growth rate ($R_g$=0.074 unit cells/s). The surface morphology (the left panel of Fig. 2(b)) is composed of nearly fully covered terraces and steps of $d_{111}$ height, indicating that the layer-by-layer mode governs regime I. The corresponding RHEED pattern confirms the atomically well-defined surface. As for the unidentified growth mode in regime II (the right panel of Fig. 2(b)), many irregular-shaped voids of one unit-cell depth are found on the terraces of the SRO surface, presumably formed by the merging of triangular voids (enclosed by white circles). It is clear that the unidentified growth mode in regime II is neither step-flow growth nor 3D island growth.[12] Although the fading of the RHEED intensity oscillations shown in regime II of Fig. 1(b) seems reminiscent of the step-flow mode observed in the growth of SRO on STO (001),[7,13] the characteristic voids and the lack of the RHEED intensity recovery after the deposition interruption suggest that the step-flow mode can be excluded as the governing growth mode in regime II. On the contrary, the higher adatom mobility at the elevated temperature of $T_g$=800 ºC and the sharper spotty RHEED pattern (the right panel of Fig. 2(b)) together suggest that the characteristic voids do not originate from a limited atomic-mobility-induced kinetic roughening. This therefore rules out the 3D island mode as a possible governing growth mode in regime II. Rather, the adatom mobility at $T_g$=800 ºC is sufficiently high that the adatoms can overcome diffusion barriers and be relocated for the surface free energy reduction, manifesting the three-fold symmetry of the (111)-oriented pseudocubic surface. This roughening behavior seems related with thermal roughening, in which surface roughening represented by creation of steps and faceting phenomena is induced by promoted step-edge barrier crossing and atom exchange with underlying terraces.[14] Other studies[6,15] also have reported that a similar three-fold symmetry can be observed in the form of many triangular islands and dents on an STO (111) surface, when the surface undergoes thermal roughening by adequately sustained high-temperature annealing.

Next, in order to investigate the effect of the growth rate on the growth mode, and to determine the possibility of growth-rate-controlled layer-by-layer growth, we changed the growth rate (by varying the pulse repetition rate) to $R_g$=0.222 unit cells/s for $T_g$=800 ºC and to $R_g$=0.019 unit cells/s for $T_g$=700 ºC. With the higher growth rate ($R_g$=0.222 unit cells/s) at $T_g$=800 ºC [Fig. 3(a)], the layer-by-layer growth mode, which was observed in regime I for $R_g$=0.074 unit cells/s with the equal temperature ($T_g$=800 ºC), is suppressed, but again prevails in regime II, bearing a resemblance to that shown in Fig. 1(a). This layer-by-layer growth mode is evidenced by small 2D islands on the terraces of the SRO surface (the right panel of Fig. 3(a)), observed after completion of growth in regime II. Due to the high supersaturation of PLD, a large number of small clusters are nucleated in the course of the duration time of a laser pulse. At such a high growth temperature as $T_g$=800 ºC, adatoms are easily detached from the clusters during the time period between the laser pulses. On a (111)-oriented pseudocubic surface, the detached mobile adatoms can migrate to form such a surface feature filled with characteristic voids, whereas the detached ones join the vicinal steps, as demonstrated in the step-flow growth mode on (001)-oriented pseudocubic surface. Such a high growth rate as $R_g$=0.222 unit cells/s, however, delivers a new batch of adatoms before the extinction of the clusters. Then, the clusters grow by accretion, coalescing to form a complete monolayer, and resulting in the 2D layer-by-layer growth mode in regime II.

By contrast, as seen in Fig. 3(b), applying the lower $R_g$=0.019 unit cells/s growth rate at $T_g$=700 ºC gives rise to the layer-by-layer growth mode in regime I, which was suppressed with the $R_g$=0.074 unit cells/s, while the RHEED intensity oscillations in regime II are damped. The surface morphology (the right panel of Fig. 3(b)), observed after completion of the growth in regime II, shows many irregular-shaped voids including triangular ones, evidencing the growth mode changes from the layer-by-layer mode, employed in regime I, to the unidentified growth mode. Such a low growth rate as $R_g$=0.019 unit cells/s prohibits the roughening in regime I, thus allowing for the layer-by-layer mode, but permits the extinction of clusters in regime II, resulting in the appearance of irregular-shaped voids.

In summary, we grew SRO thin films on atomically flat single-terminated STO (111) substrates. We found 3D island mode, layer-by-layer mode, and a certain unidentified mode in different thickness ranges, as well as growth mode transitions between them, which can be controlled by varying the growth rate and temperature parameters. These growth parameters can affect kinetic parameters such as adatom mobility and flux, which can, in turn, determine growth modes. Controllable growth parameters allow for layer-by-layer growth of SRO over a wide thickness range as well as for atomic-level control of atomically sharp heterointerfaces.

## Acknowledgements

We thank Profs. Cheol Seong Hwang and Miyoung Kim for the valuable discussions. This work was supported by Creative Research Initiatives (ReC-SDSW) of MEST/KOSEF.

**Figure captions**

Fig. 1. (Color online) RHEED intensity profiles during growth of SRO films on atomically flat single-terminated STO (111) substrates at $T_g=700$ ºC and $R_g=0.074$ unit cells/s in (a), and at $T_g=800$ ºC and $R_g=0.074$ unit cells/s in (b).

Fig. 2. (Color online) AFM surface morphology images and corresponding RHEED patterns taken after deposition interruption within regimes I and II for $T_g=700$ ºC in (a), and for $T_g=800$ ºC in (b).

Fig. 3. (Color online) RHEED intensity profiles during growth of SRO films on single-terminated STO (111) substrates at $T_g=800$ ºC and $R_g=0.222$ unit cells/s in (a) and $T_g=700$ ºC and $R_g=0.019$ unit cells/s in (b). The right panels show the corresponding AFM images taken after the deposition interruption at the stage indicated by the closed arrow in regime II, followed by the saturation of the RHEED intensities. The white box is a zoom-in image of the area indicated by the smaller white ellipse.

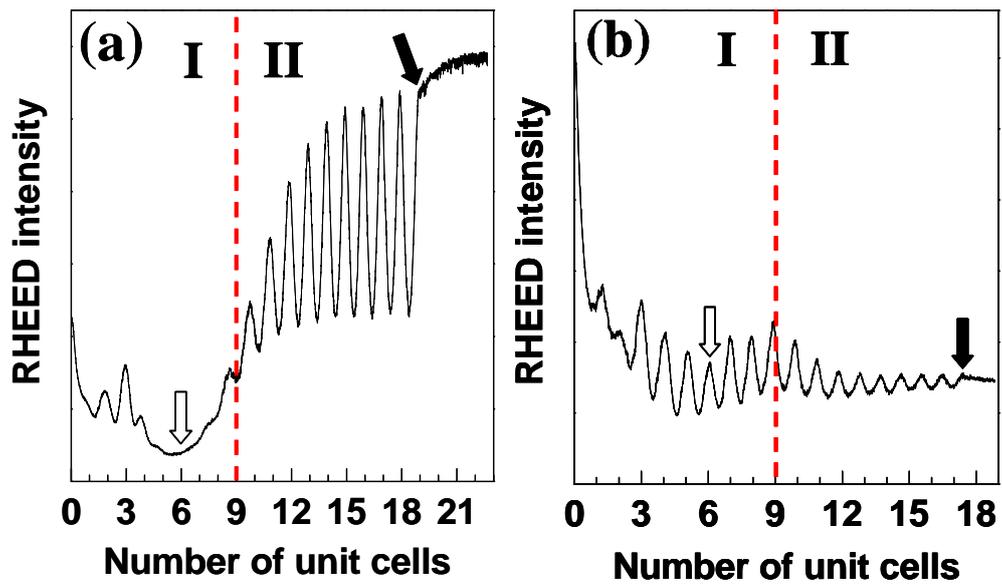

**Fig. 1**

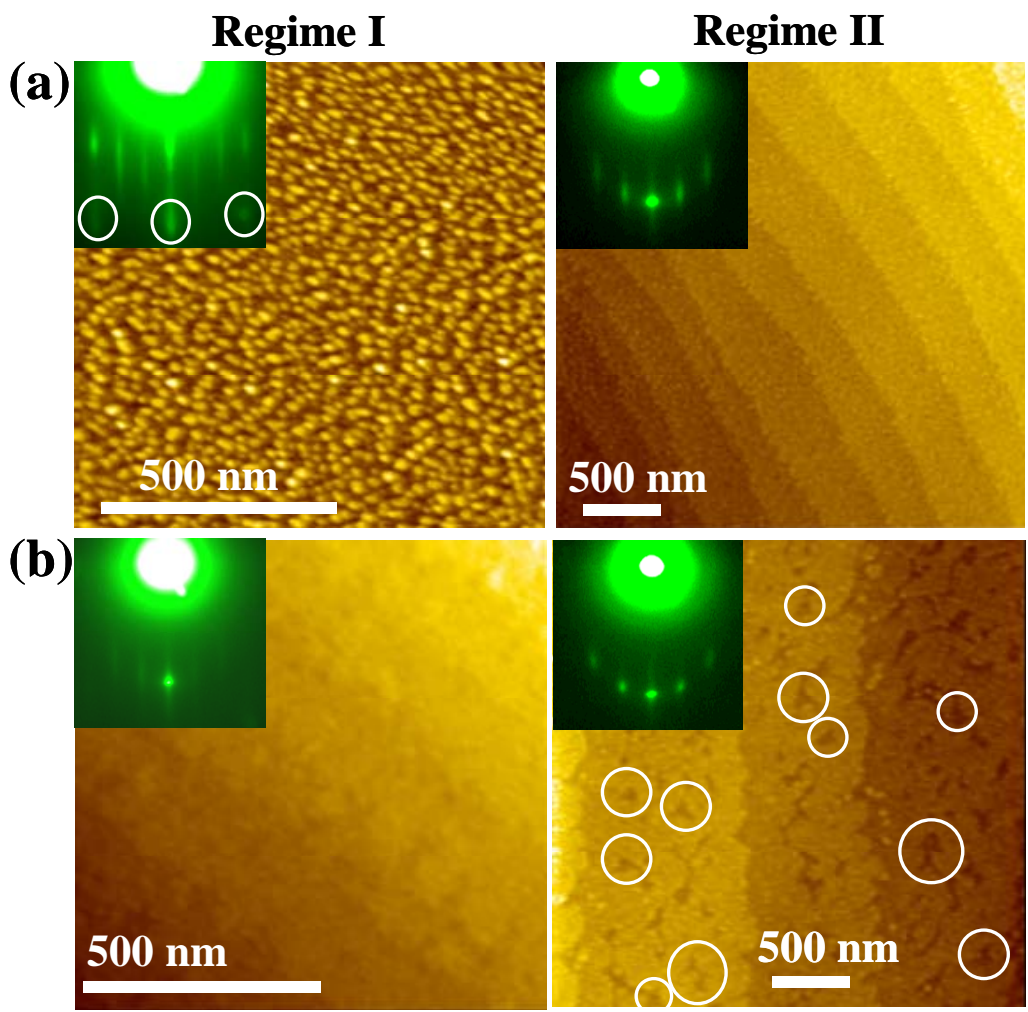

Fig. 2

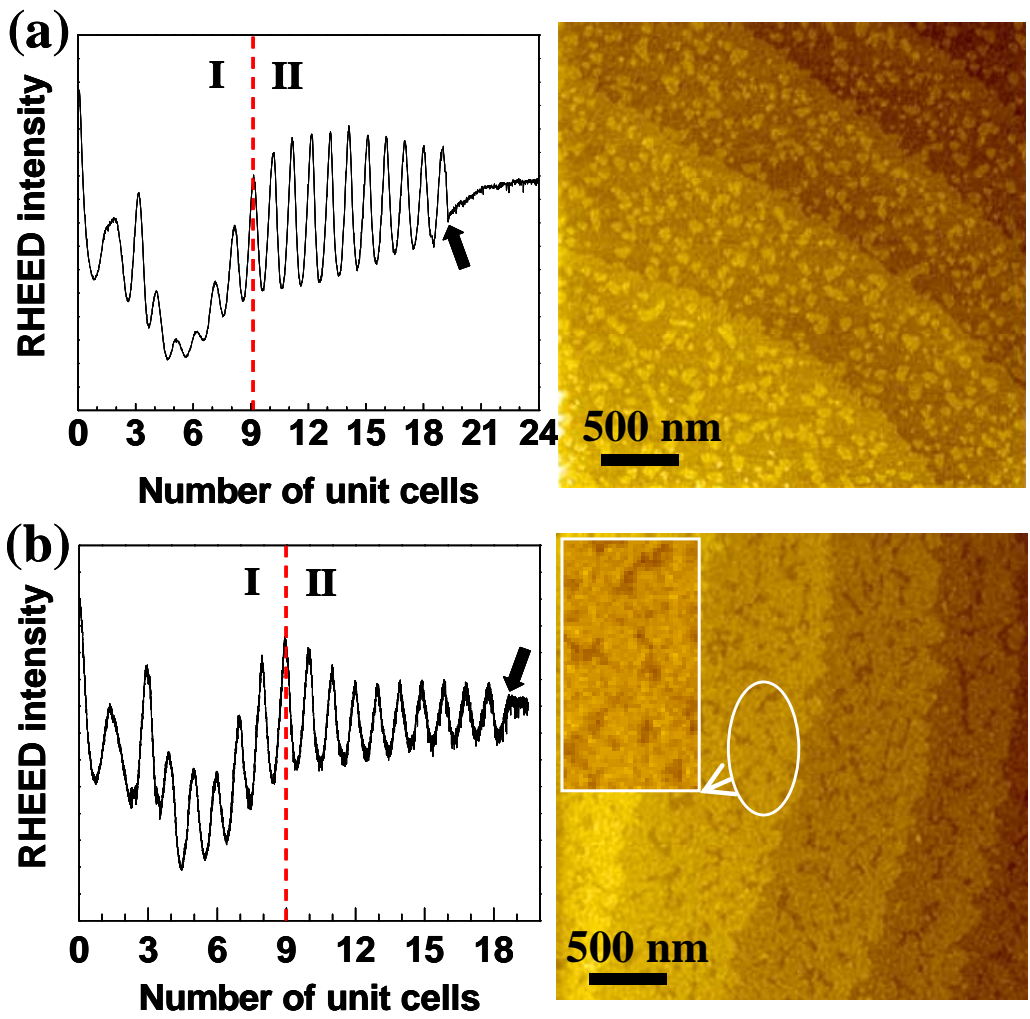

Fig. 3